\documentclass[preprint,a4paper,twoside,longnamesfirst]{aastex}

\newcommand{\COtwozero}{\mbox{$\mathrm{CO\:2\!-\!0}$}}
\newcommand{\COtwoone}{\mbox{$\mathrm{CO\:J=2\!-\!1}$}}
\newcommand{\Htwo}{\mbox{$\mathrm{H_2}\:v\!=\!1\!-\!0\:\mathrm{S(1)}$}}
\newcommand{\Brg}{Brackett~$\gamma$}
\newcommand{\kms}{\,\mathrm{km\,s^{-1}}}
\newcommand{\Hz}{\,\mathrm{Hz}}

\newcommand{\GHz}{\,\mathrm{GHz}}

\newcommand{\cm}{\,\mathrm{cm}}
\newcommand{\pc}{\,\mathrm{pc}}
\newcommand{\kpc}{\,\mathrm{kpc}}
\newcommand{\Mpc}{\,\mathrm{Mpc}}
\newcommand{\Msol}{\,{\cal M}_{\sun}}
\newcommand{\Mdyn}{{\cal M}_{dyn}}
\newcommand{\Lsol}{\,\mathrm{L_\sun}}
\newcommand{\Lbol}{L_{\mathrm{bol}}}
\newcommand{\W}{\,\mathrm{W}}

\begin{document}

\title{Stellar Dynamics and the implications on the merger evolution
in \objectname[]{NGC~6240}}
\author{M.~Tecza, R.~Genzel, L.~J.~Tacconi, S.~Anders,
L.~E.~Tacconi--Garman and N.~Thatte}
\affil{Max-Planck-Institut f\"ur extraterrestrische Physik (MPE)}
\affil{D-85748 Garching, Germany}

\slugcomment{Submitted to ApJ: 9 November 1999; Revised: 24 January 2000}

\begin{abstract}
We report near-infrared integral field spectroscopy of the luminous
merging galaxy \objectname[]{NGC~6240}.  Stellar velocities show that
the two K-band peaks separated by $1\farcs6$ are the central parts of
inclined, rotating disk galaxies with equal mass bulges.  The
dynamical masses of the nuclei are much larger than the stellar mass
derived from the K-band light, implying that the progenitor galaxies
were galaxies with massive bulges.  The K-band light is dominated by
red supergiants formed in the two nuclei in starbursts, triggered
$\approx2\times10^7$ years ago, possibly by the most recent
perigalactic approach.  Strong feedback effects of a superwind and
supernovae are responsible for a short duration burst
($\approx5\times10^6$ years) which is already decaying.  The two
galaxies form a prograde-retrograde rotating system and from the
stellar velocity field it seems that one of the two interacting
galaxies is subject to a prograde encounter.  Between the stellar
nuclei is a prominent peak of molecular gas ($\mathrm{H_2}$, CO).  The
stellar velocity dispersion peaks there indicating that the gas has
formed a local, self-gravitating concentration decoupled from the
stellar gravitational potential.  \objectname[]{NGC~6240} has
previously been reported to fit the paradigm of an elliptical galaxy
formed through the merger of two galaxies.  This was based on the
near-infrared light distribution which follows a $r^{1/4}$-law.  Our
data cast strong doubt on this conclusion: the system is by far not
relaxed, rotation plays an important role, as does self-gravitating
gas, and the near-infrared light is dominated by young stars.
\end{abstract}


\section{Introduction}
The luminous infrared galaxy (LIRG) \objectname[]{NGC~6240}
\citep{WRI1984,THR1990} has a remarkable disturbed morphology which is
characterised in the visible by loops, branches and arms extending out
to $50\kpc$ \citep{FOS1979}.  From this large-scale morphology, and
the discovery of two nuclei in the central region \citet{FRI1983}
concluded that \objectname[]{NGC~6240} is an interacting and merging
system of two galaxies.  This conclusion is supported by spectroscopic
observations in the visible \citep{FOS1979,FRI1985} and near-infrared
wavelength range \citep{HER1990,VWE1993,SUG1997}.
\objectname[]{NGC~6240} is the most luminous source of near-infrared
line emission ($L(\mathrm{H_2})_{\mathrm{tot}}\approx10^9\Lsol$,
\citet{WRI1984}).  The $\mathrm{H_2}$ emission is most likely excited
in shocks triggered by the collision of the two galaxies.

\objectname[]{NGC~6240} has a bolometric luminosity $\Lbol\approx
L_{\mathrm{IR}}=6\times10^{11}\Lsol$ (for
$\mathrm{H_0}=75\kms\Mpc^{-1}$, $\mathrm{D=97\Mpc}$). This bolometric
luminosity classifies \objectname[]{NGC~6240} almost as an ultra
luminous infrared galaxy (ULIRG; $L_{\mathrm{IR}}\ge10^{12}\Lsol$).
LIRGs and ULIRGs emit more than 90\% of their bolometric luminosity in
the infrared wavelength range.  There is consensus that the infrared
emission comes from warm dust but what is the primary energy source
responsible for heating the dust is still a matter of debate.  The two
possibilities are a deeply dust embedded active galactic nucleus (AGN)
or a super starburst.  Observations with the Infrared Space
Observatory (ISO) \citep{GEN1998} show that the majority of ULIRGs,
including \objectname[]{NGC~6240}, are starburst dominated.  On the
other hand, detection of a highly absorbed hard X-ray source indicates
that \objectname[]{NGC~6240} also contains a powerful AGN
\citep{VIG1999}.  Most ULIRGs are interacting or merging systems with
the star formation and/or AGN activity triggered by gas compression
and inflows towards the center \citep{SAN1988}.  As such
\objectname[]{NGC~6240} is a nearby test case of a luminous merger.
It can be used to test predictions of interacting and merging galaxies
\citep{BAR1996,MIH1996}.  The effect of the interaction on the star
formation can be compared with model results to assess the true nature
of ULIRGs.  To investigate the stellar content and the properties of
the starburst of \objectname[]{NGC~6240}, to determine the stellar
dynamics and to test predictions of interacting galaxy models we have
carried out high resolution near-infrared spectroscopy with the
Max-Planck-Institut f\"ur extraterrestrische Physik (MPE) integral
field spectrometer 3D.

\section{Observations and Data Reduction}
\objectname[]{NGC~6240} was observed with the MPE near-infrared
imaging spectrograph 3D \citep{WEI1996} in conjunction with the
tip-tilt correction adaptive optics system ROGUE \citep{THA1995} in
two observing runs.  3D is an integral field spectrograph that
simultaneously obtains spectra for each of 256 spatial pixels covering
a square field of view with over 95\% fill factor.  In both observing
runs the spectral resolving power
($\mathrm{R}\equiv\lambda/\Delta\lambda$) was 2000 and Nyquist sampled
using two settings of a piezo-driven flat mirror.

The first observing run took place in April 1996 at the ESO
$2.2\,\mathrm{m}$ telescope on La~Silla, Chile.  The pixel scale was
$0\farcs3$ per pixel, and a wavelength range from $2.18\micron$ to
$2.45\micron$ was covered.  The total on-source integration time was
$4600\,\mathrm{s}$ with individual frame integration times of
$300\,\mathrm{s}$ or $400\,\mathrm{s}$.  The same amount of time was
spent off-source $1\arcmin$~E and W of the nuclear region of
\objectname[]{NGC~6240} for sky background subtraction.  The seeing
during the observations was better than $0\farcs8$.

The second observing run took place in March 1998 at the Anglo
Australian Telescope in Coonabarabaran, Australia.  Here the pixel
scale was $0\farcs4$ per pixel and a wavelength range from
$2.155\micron$ to $2.42\micron$ was covered.  This wavelength setting
includes the \Htwo\ emission line at a redshifted wavelength of
$2.173\micron$.  The total on-source integration time was
$3100\,\mathrm{s}$ with individual integrations of $100\,\mathrm{s}$
each.  Again, the same amount of time was spent off-source
$1\arcmin$~E of the nuclear region of \objectname[]{NGC~6240} to
subtract the sky background.  The seeing throughout the observations
was better than $1\arcsec$.

To correct for the atmospheric transmission a reference star was
observed before and after the science integrations.  The data were
reduced using the set of 3D data analysis routines written for the
GIPSY \citep{VHU1992} data reduction package.  This included
wavelength calibration, spectral and spatial flat fielding, dead and
hot pixel correction, and division by the reference stellar spectrum.
Data cubes from the individual exposures were recentered and added
using the centroid of the broadband emission from the southern
nucleus.  Absolute flux calibration was established by comparison of
the broadband emission with K-band photometry by \citet{THR1990}.
Emission line and absorption line maps as well as the continuum map
were extracted by performing a linear fit through line-free regions in
the vicinity of the lines in the spectrum of each spatial pixel.

\section{Results}
\subsection{Line Maps and Spectra}
The spectrum of \objectname[]{NGC~6240} in the range from
$2.15\micron$ to $2.45\micron$ contains a number of molecular and
atomic emission lines, as well as stellar absorption features.

\begin{figure}
\includegraphics[width=\columnwidth]{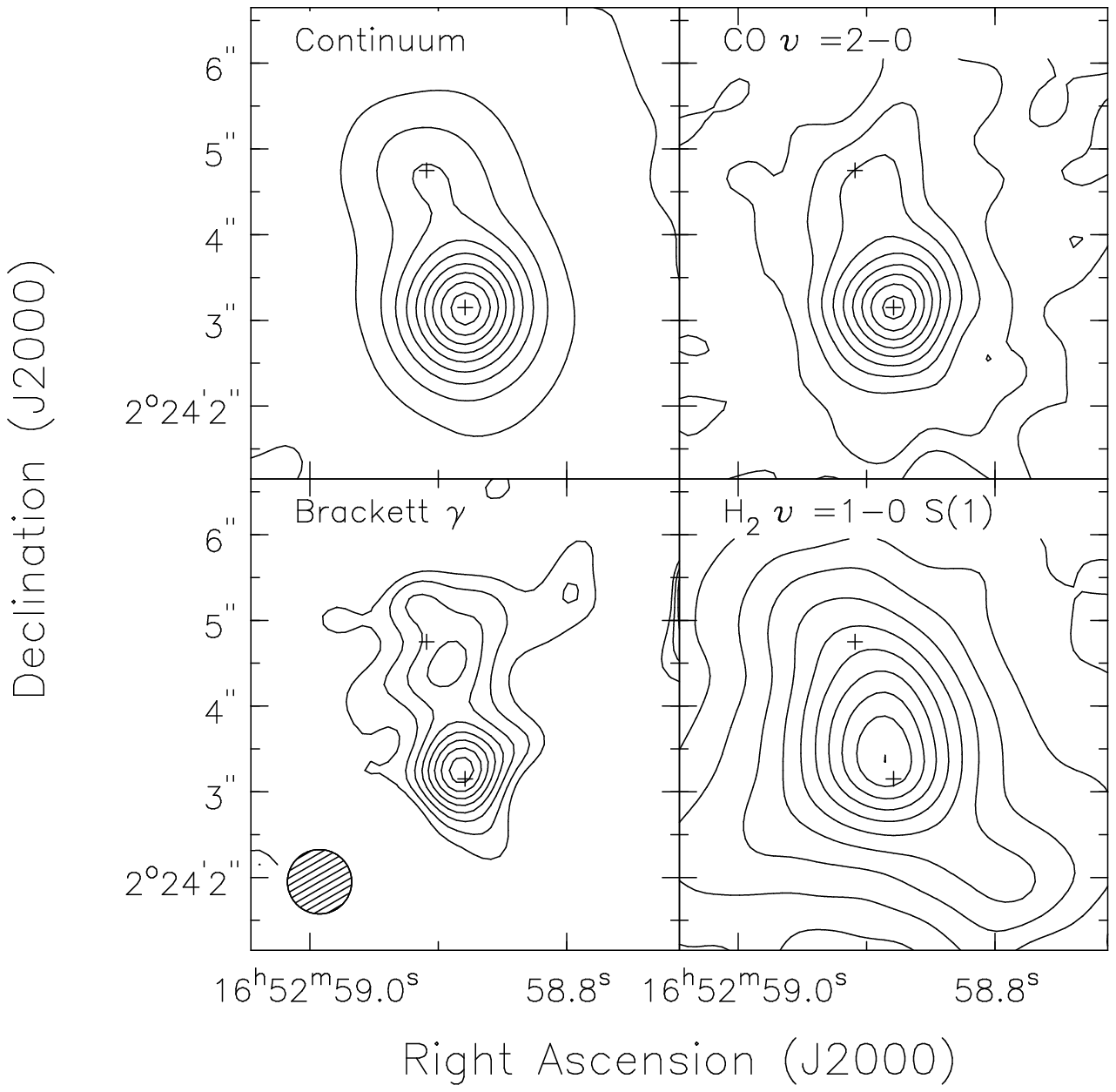}
\figcaption{
Selected line maps of NGC~6240.  The crosses in all four panels denote
the position of the K-band nuclei.  Contours are in 10\% steps.
Top-left: Integrated continuum emission in the wavelength range from
$2.2\micron$ to $2.4\micron$.  Top-right: Map of the \COtwozero\
absorption bandhead depth.  Bottom-left: \Brg\ emission line map.
Bottom-right: Integrated \Htwo\ emission line map.
\label{Fig:Linemaps}}
\end{figure}

Figure \ref{Fig:Linemaps} shows the line-free continuum map in the
wavelength range from $2.2\micron$ to $2.45\micron$, a \Brg\ map and a
map of the \COtwozero\ absorption bandhead.  All three maps are
dominated by two compact sources (hereafter referred to as nuclei)
separated by $1\farcs6$.  The double nuclear structure has been
observed previously in the visible (B, V, R) \citep{KEE1990} and NIR
(J, H, K) bands \citep{THR1990} but the maps are also similar to what
is observed at other wavelengths \citep{CON1982,COL1994,KET1997}.  In
addition to the double nucleus the \Brg\ map shows emission extended
out to $1\farcs5$ northwest of the northern nucleus in the direction
of the third radio emission peak N3 observed by \citet{COL1994}.

Also shown in Figure \ref{Fig:Linemaps} is the integrated \Htwo\
emission.  In contrast to the K-band continuum the \Htwo\ emission
shows a single emission peak \emph{between} the nuclei with filaments
extending from it.  The emission peak is at a separation of
$0\farcs35$ from the southern nucleus towards the northern nucleus and
is extended in the northern direction. This $\mathrm{H_2}$ morphology
has been previously observed with Fabry-Perot imaging techniques by
\citet{HER1990}, \citet{VWE1993} and \citet{SUG1997}.  The
$\mathrm{H_2}$ morphology is very similar to the \COtwoone\
morphology, indicating that also the near-infrared $\mathrm{H_2}$
emission lines trace the bulk of the molecular material in
\objectname[]{NGC~6240}.

\begin{figure}
\includegraphics[width=\columnwidth]{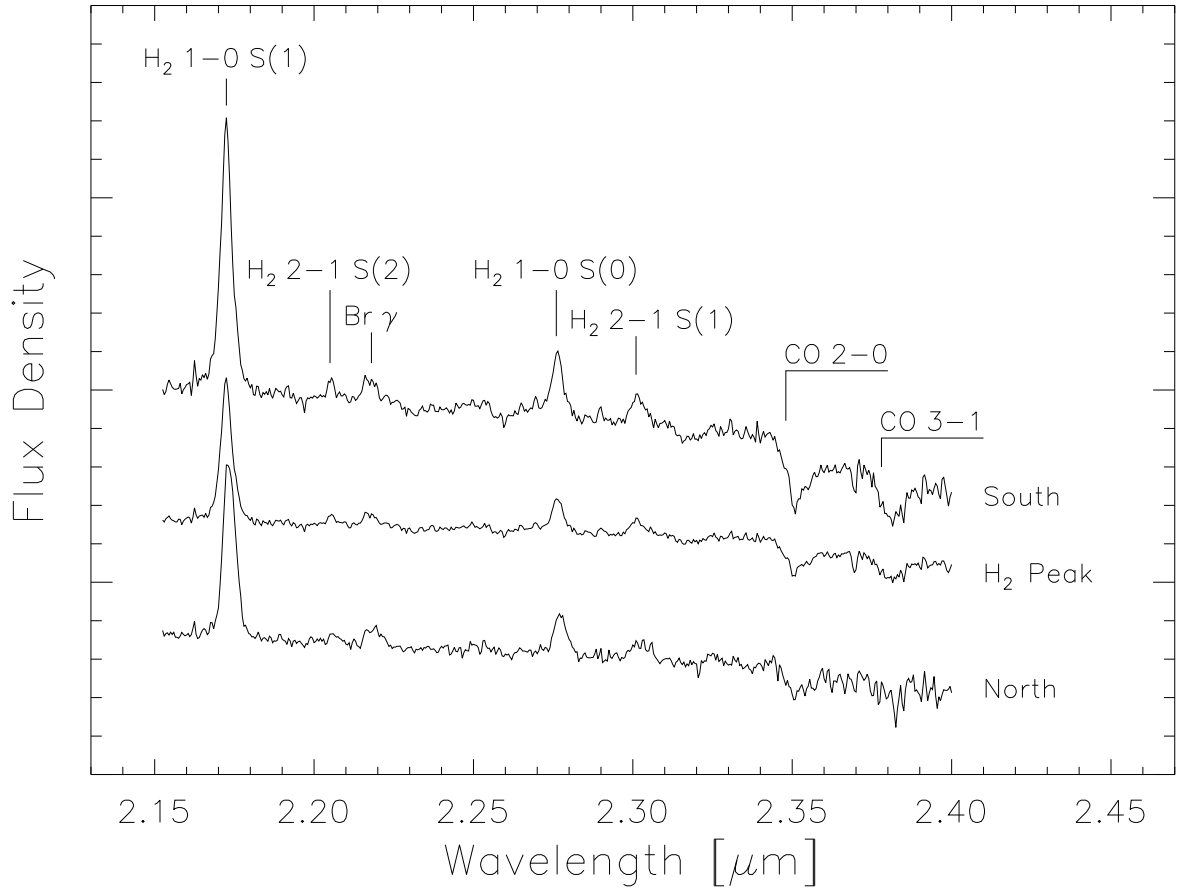}
\figcaption{
$R=2000$ spectra of the \Htwo\ emission peak, the northern nucleus and
southern nucleus of NGC~6240 (each in an $1\arcsec$ diameter circular
aperture).
\label{Fig:Spectra}}
\end{figure}

The line maps in Figure \ref{Fig:Linemaps} show three major regions of
interest.  These are the two nuclei, and the region of the \Htwo\
emission peak between them.  Figure \ref{Fig:Spectra} shows the
spectra of these regions, each within an $1\arcsec$ diameter circular
aperture.  Even though the peak of the $\mathrm{H_2}$ emission lies
between the nuclei, $\mathrm{H_2}$ dominates the line emission in all
three regions.  In fact, \objectname[]{NGC~6240} has the highest ratio
of \Htwo\ luminosity to bolometric luminosity of any galaxy which has
thus far been observed \citep{WRI1984}.  The $\mathrm{H_2}$ lines are
very broad and have a full width half maximum (FWHM) of $550\kms$ with
line wings extending over $1600\kms$ full width zero power (FWZP).
The emission line profiles are asymmetric with red and/or blue wings
whose strength varies over the emission region. Figure \ref{Fig:H2&CO
Profiles} shows a comparison of the line-of-sight velocity profiles of
the \Htwo\ and \COtwoone\ lines for the nuclear region of
\objectname[]{NGC~6240}.

A characteristic of \objectname[]{NGC~6240} is its low \Brg\
equivalent width \citep[e.g.][]{LES1988,VWE1993}.  In earlier work
\citep[see references in][]{VWE1993} the line was not or only
marginally detected.  In the improved sensitivity 3D data the line is
now clearly detected.  The \Brg\ equivalent width of the northern and
southern nucleus is $6.5\pm1.6\mathrm{\AA}$ and
$2.9\pm0.4\mathrm{\AA}$, respectively.  A line map and spectra in
different positions can be extracted from the data cube.  The FWHM of
the \Brg\ line is $716\kms$ and $689\kms$ for the northern and
southern nucleus, respectively, measured over a $1\arcsec$ diameter
circular aperture.  The northern nucleus is redshifted with respect to
the southern nucleus by $137\pm49\kms$, a value similar to the
velocity difference of $147\kms$ measured by \citet{FRI1985} with
optical emission and absorption lines.

The 3D spectra also show that \objectname[]{NGC~6240} has deep CO
absorption bandheads with a large velocity dispersion.  From similar
data \citet{LES1994} and \citet{DOY1994} deduce a velocity dispersion
of $350\kms$ and $359\kms$, respectively, for the region of
\objectname[]{NGC~6240} containing both nuclei.  Using 3D, we are able
to resolve spatially the velocity dispersions of each nucleus.  A
detailed analysis of this is presented in \S\ref{Sec:Kinematics}.  The
equivalent width of the \COtwozero\ bandhead is determined using the
wavelength intervals given in \citet{ORI1993} and corrected for the
velocity broadening using the formula in \citet{OLI1995}.  The
velocity dispersions are the ones derived in \S\ref{Sec:Kinematics}
and the resulting equivalent widths are $13\pm3\mathrm{\AA}$ and
$15\pm1\mathrm{\AA}$ for the northern and southern nucleus,
respectively.

\subsection{Extinction}
\label{Extinction}
From the spectra in Figure \ref{Fig:Spectra} it is apparent that the
spectrum of the $\mathrm{H_2}$ emission peak has a shallower, i.e.\
redder, slope than the two nuclei.  Assuming an intrinsically constant
continuum slope over the entire nuclear region of
\objectname[]{NGC~6240}, this reddening can either be due to
extinction or a nonstellar contribution to the spectrum at long
wavelengths.  The spectral energy distribution (SED) of
\objectname[]{NGC~6240} \citep{DRA1990} shows no nonstellar continuum
below $5\micron$.  Above $5\micron$ thermal emission from dust with a
temperature of $\approx200\,\mathrm{K}$ is dominant.  Different
spectral slopes over the nuclear region of \objectname[]{NGC~6240}
thus are almost certainly due to extinction variations.

To derive the extinction values the spectral slope of all pixels in
the data cube are compared with the spectral slope of a K4.5
supergiant, which was observed with 3D during the La~Silla 1996
observing run.  A direct comparison of stellar spectra with the
spectrum of \objectname[]{NGC~6240} yields a K4.5 supergiant as the
best fitting template (see \S\ref{Sec:Continuum}).  With the
interstellar extinction law from \citet{DRA1989}
\begin{displaymath}
\frac{\mathrm{A_\lambda}}{\mathrm{A_V}}=0.351\cdot 
   \lambda^{-1.75}_{\mathrm{\mu m}}
\end{displaymath}
and an uniform foreground screen model (UFS) we derive the extinction
map shown in Figure \ref{Fig:Extinction}.  The extinction towards the
southern and northern nucleus is $\mathrm{A_V^S}=5.8$ and
$\mathrm{A_V^N}=1.6$, respectively.  The peak extinction is
$\mathrm{A_V}=7.2\pm0.7$.  For a mixed model, where absorbing dust and
stellar emission are completely mixed, the peak value is
$\mathrm{A_V}=18.4^{+3.1}_{-1.9}$.  The morphology of the extinction
map with a single peak between the continuum nuclei indicates a dust
concentration between the nuclei and is coincident with the \Htwo\ and
\COtwoone\ emission peaks.

\section{The Infrared Nuclei}
\subsection{The Nature of the K-band Continuum}
\label{Sec:Continuum}
To determine the stellar type dominating the near-infrared light in
starburst galaxies the \COtwozero\ absorption bandhead is an often
used spectral absorption feature.  For late-type giants and
supergiants this absorption feature is rather deep, and even with data
of medium spectral resolution the equivalent width can be used to
determine the stellar type.  There is an ambiguity in the \COtwozero\
equivalent width between KI and MIII stars, however.  From the
\COtwozero\ absorption bandhead equivalent width the spectrum of
\objectname[]{NGC~6240} can be explained either by a population of red
giants or red supergiants.  However, because of the high spectral
resolution of our data, we can resolve the discrepancy by a direct
comparison of the spectrum of \objectname[]{NGC~6240} with both
stellar types.  Figure \ref{Fig:Supergiants} compares the spectrum of
the CO absorption bandheads of the southern nucleus of
\objectname[]{NGC~6240} with four stellar spectra of giants and
supergiants also observed with 3D at the same spectral resolution
\citep{SRE1999}.  For the comparison the spectrum of
\objectname[]{NGC~6240} was shifted into the restframe of zero
redshift, and the stellar spectra were broadened with a Gaussian
velocity profile of $600\kms$ FWHM.  Clearly the K and M giant spectra
do not fit the deep absorption features and the continuum slope.  On
the other hand the absorption bandheads of a M3+ supergiant are too
deep to represent the \objectname[]{NGC~6240} spectrum.  The best
matching stellar spectrum is that of a K4.5 supergiant.  However, its
CO absorption bandheads are still somewhat too shallow and we conclude
that the near-infrared light is due to late K (K5) or early M (M0/M1)
supergiants.  Our conclusions are consistent with those of
\citet{SUG1997} whose data have lower spectral resolution but cover a
wider spectral range.  The fact that the K-band light of the nuclei is
dominated by red supergiants indicates a starburst as the source of
the K-band luminosity.  The presence of supergiants implys that the
starburst was triggered quite recently.

\subsection{Starburst Simulations}
\label{Sec:Starburst}
To further constrain the age and other characteristics of the
starburst we used the program STARS \citep{SRE1999} to simulate
several starburst scenarios.  In Figure \ref{Fig:Starburst} we show
model predictions for the variations of \Brg\ equivalent width, the
K-band luminosity to stellar mass ratio ($L_{\mathrm{K}}/{\cal M_*}$)
and the starburst bolometric luminosity to K-band luminosity ratio
($L_{\mathrm{bol}}^*/L_{\mathrm{K}}$) for four burst durations: $1$,
$5$ and $20$ million years as well as continuous star formation.  We
adopted a \citet{SAL1955} initial mass function (IMF) for masses
between $100\Msol$ and $1\Msol$.  The calculated luminosity output of
the starburst is insensitive to the shape of the IMF at the low mass
end ($<1\Msol$).  The low mass stars do contribute to the total mass
of the stars formed in the starburst, however.  To take this into
account, we integrated a Miller-Scalo IMF \citep{MIL1979} from
$1\Msol$ down to $0.08\Msol$, the lower mass limit for hydrogen
burning.  Because stars with masses $>25\Msol$ do not evolve into red
supergiants, the K-band luminosity does not depend on the choice of
the upper mass cutoff, and due to their small number, their
contribution to the total mass is negligible.

\subsection{Starburst Age and Duration}
The \COtwozero\ absorption bandhead equivalent width (or better the
presence of late type supergiants) allows us to determine the age of
the starbursts in \objectname[]{NGC~6240}.  The late K or early M
supergiants that dominate the K-band light of the nuclei have masses
from $10$ to $20\Msol$ and a typical age of $15$ to $25$ million
years.  Since red supergiants dominate the K-band luminosity only
during this period, the starburst must be of similar age.  This age is
shown in Figure \ref{Fig:Starburst} as a vertical hatched bar.

The \Brg\ emission line equivalent width allows us to constrain the
duration of the starburst activity.  The agreement of the \Brg\ and
K-band continuum morphologies indicates that both the line and
continuum emission originates in the same star forming regions.  The
$\mathrm{H_2}$ emission can be explained by excitation in slow
C-shocks \citep{VWE1993,SUG1997,EGA1998}.  Since the morphology of the
$\mathrm{H_2}$ emission differs from that of the \Brg\ emission,
shocks most likely do not contribute to the \Brg\ emission.  In a
starburst, only stars with masses $>20-30\Msol$ ionise their
surrounding medium to produce \Brg\ emission, but due to their high
mass their lifetime is very short ($<10$ million years).  The low
\Brg\ equivalent width in \objectname[]{NGC~6240} indicates a low
number of hot, young stars, either because the starburst is aging or
because stars with masses $>20\Msol$ were never formed.  We prefer the
former explanation, because of the signatures for $50$ to $100\Msol$
stars in nearby starburst templates \citep[see][]{THO1999}.

The middle panel of Figure \ref{Fig:Starburst} shows that the
starburst duration is $<5$ million years.  The short starburst
duration can be explained by strong negative feedback effects from the
starburst itself.  The onset of vigorous star formation produces
young, massive stars and, thus, supernovae in a very short time
($\lesssim5$ million years).  The winds of these stars deplete the
molecular gas and the star formation subsides.  The existence of a
superwind in \objectname[]{NGC~6240} was shown by \citet{HEC1990} from
$\mathrm{H\alpha}$ line mapping and spectroscopy.

The duration of the starbursts is much smaller than the age of the
starburst but comparable to the dynamical time scale of $\approx7$
million years of the two rotating nuclei (see \S\ref{Sec:Rotation}).
The time scale for the interaction and merging of the two galaxies is
several hundred million years and hence much larger than the starburst
duration \emph{and} starburst age.  Such short but violent star
formation events are predicted by models of interacting and merging
galaxies, where they are triggered by the close encounters of the
interacting partners \citep{MIH1996}.

\subsection{Stellar Mass in the Starburst}
The light-to-mass ratio ($L_{\mathrm{K}}/{\cal M}_*$) of the starburst
can be calculated from simulations of the K-band luminosity and the
total mass of stars formed in the starburst.  The middle panel of
Figure \ref{Fig:Starburst} shows $L_{\mathrm{K}}/{\cal M_*}$ as a
function of starburst age.  It peaks at the time when the red
supergiants dominate the K-band luminosity.  For the starburst age in
\objectname[]{NGC~6240} the simulation yields $1\le
L_{\mathrm{K}}/{\cal M}_*\le 3$.  From the dereddened K-band
luminosities in Table \ref{Tab:Luminosity} within $1\arcsec$ a stellar
mass ${\cal M}_*=0.4-1.2\times10^8\Msol$ and ${\cal
M}_*=0.8-2.3\times10^8\Msol$ is derived for the northern and southern
nucleus, respectively.  Here the lower mass cutoff was assumed to be
$0.08\Msol$.  If the lower mass cutoff was $0.25\Msol$, $0.5\Msol$ or
$1\Msol$ the estimated stellar mass would be smaller by $12\%$, $23\%$
or $39\%$, respectively.  This is the mass of young stars formed in
the starburst contained within the central $500\pc$ of the two nuclei.
It shows that the two nuclei are massive objects and possibly could be
the nuclei of two galaxies involved in a collision, based on the mass
of young stars alone.

\subsection{The Luminosity of the Nuclei}
The bottom panel in Figure \ref{Fig:Starburst} shows the starburst
bolometric luminosity to K-band luminosity ratio
($L_{\mathrm{bol}}^*/L_{\mathrm{K}}$) as a function of starburst age.
For a starburst age of $15$ to $25$ million years this ratio is
$\approx100$.  From the dereddened K-band luminosity within $2\arcsec$
diameter of each nucleus (see Table \ref{Tab:Luminosity}) this ratio
yields $\Lbol^*\approx0.7\times10^{11}\Lsol$ and
$\Lbol^*\approx1.7\times10^{11}\Lsol$ for the northern and southern
nucleus, respectively, totaling to
$\Lbol^*\approx2.4\times10^{11}\Lsol$.  This is \onethird\ to
\onehalf\ of the entire bolometric luminosity
$\Lbol\approx6\times10^{11}\Lsol$ of \objectname[]{NGC~6240} in the
IRAS beam.  Within a $5\arcsec$ diameter aperture the total K-band
luminosity $L_{\mathrm{K}}=6.1\times10^9\Lsol$ is
$\approx$\slantfrac{1}{100} of the total bolometric luminosity.  In
the bottom panel of Figure \ref{Fig:Starburst} this ratio is shown as
a horizontal, hatched bar.  Within the errors of the simulations the
entire bolometric luminosity of \objectname[]{NGC~6240} can be
explained by the starburst.

The claim that the starbursts in the nuclei contribute significantly
to the bolometric luminosity of \objectname[]{NGC~6240} is further
strengthened by radio observations.  The radio flux and the far
infrared (FIR) luminosity of \objectname[]{NGC~6240} follow the
radio-to-FIR correlation for starburst galaxies.  \citet{LIS1996}
determined empirically the parameter
\begin{displaymath}
\mathrm{q_{2.4\GHz}}= \log\left(\frac{S_{\mathrm{FIR}}}{3.75\times
10^{12}\cdot S_{\mathrm{2.4\GHz}}}\right)
\end{displaymath}
relating the radio-power to the FIR-luminosity.  For starburst
galaxies $\mathrm{q_{2.4\GHz}}=2.40\pm0.22$.  Taking the radio-power
$S_{2.4\GHz}=1.7\times10^{23}\W\Hz^{-1}$ at $\nu=2.4\GHz$ for the
nuclear region including both nuclei, and comparing it with the
FIR-luminosity $S_{\mathrm{FIR}}=1.9-2.3\times10^{38}\W$ yields for
\objectname[]{NGC~6240} $\mathrm{q_{2.4\GHz}}=2.47-2.55$, within the
uncertainties of the radio-to-FIR correlation.  If we include the
radio-emission from the western region \citep[region W in][]{COL1994},
which has no counterpart in the visible or infrared, we get
$\mathrm{q_{2.4\GHz}}=2.31-2.39$, also within the uncertainties of the
q-parameter for starburst galaxies.

X-ray observations and the detection of $25.9\micron$ [\ion{O}{4}]
emission indicate the presence of a powerful AGN in
\objectname[]{NGC~6240} as well
\citep{SUL1998,KOM1998,IWA1998,GEN1998,VIG1999}.  In fact,
\citet{VIG1999} conclude from their \emph{Beppo}SAX data that the
highly absorbed ($\mathrm{N_H}\approx2\times10^{24}\cm^{-2}$) AGN can
account for most of the bolometric luminosity of
\objectname[]{NGC~6240}.  These two conclusions seem to contradict
each other .  However, extrapolating from the near/mid-infrared
spectroscopy and hard X-ray photometry to bolometric luminosity and
the possibility of significant correction of the observed luminosity
for anisotropy introduce uncertainties of factors $\ge2$.  We conclude
that \objectname[]{NGC~6240} is a composite object where both star
formation and AGN play a role.

At the position of the K-band nuclei also \Brg\ emission is observed,
however, very little continuum emission is detected at the position of
the north-west extension of the \Brg\ emission.  This could be an
indication that the \Brg\ emission is arising from a very young
starburst that has not yet formed enough stars to contribute to the
continuum emission, but whose hot, young stars have already ionised
the interstellar medium (ISM).  If true, we can estimate the
bolometric luminosity of this emission region using the radio-to-FIR
correlation and get $S_{\mathrm{FIR}}=0.4\times10^{11}\Lsol$.  This is
about half the bolometric luminosity of the northern nucleus and
$\approx15\%$ of the total bolometric luminosity of
\objectname[]{NGC~6240}, a non-negligible contribution to the total
bolometric luminosity of \objectname[]{NGC~6240}.  \citet{COL1994}
have interpreted their third radio peak N3, based on its steeper radio
spectrum, as a clump of electrons, driven away from the nucleus by a
superwind.  From the \Brg\ emission we think this explanation is less
likely and favour instead that N3 is a young starburst region.

\section{Stellar Kinematics}
\label{Sec:Kinematics}
In the near-infrared wavelength range from $2-2.5\micron$ stellar
kinematics can be determined from absorption features of Ca, Na and
CO.  By far the strongest absorption feature is the \COtwozero\
absorption bandhead at $2.29\micron$.  It has a very sharp blue edge
which is very sensitive to stellar motions.  Another advantage is that
the extinction in the K-band is only $10\%$ of the visual extinction.
The \COtwozero\ absorption bandhead is, therefore, very well suited
for the determination of the stellar kinematics in the gas- and
dust-rich infrared galaxies.

We have derived the velocity dispersions and radial velocities of the
nuclei of \objectname[]{NGC~6240} by using a modified Fourier
correlation quotient (FCQ) method \citep{BEN1990,AND1999} and a direct
fitting method on the \COtwozero\ absorption bandhead data.  In the
FCQ method the cross-correlation of the template spectrum with the
galaxy spectrum and the auto-correlation of the template spectrum are
computed in the Fourier domain.  Then the quotient of the
cross-correlation and the auto-correlation is calculated.  To suppress
high-frequency noise a Wiener filter is applied to the
correlation-quotient.  Fourier transformation of the correlation
quotient finally yields the line-of-sight velocity distribution in the
galaxy spectrum.  In the direct fitting method a stellar template
spectrum was convolved with a Gaussian broadening function.  The
broadening function was parameterised by a radial velocity and a
velocity dispersion.  With a least-$\chi^2$ fitting technique the best
fitting set of parameters was computed.  To derive the kinematic
parameters with the FCQ method the deconvolved velocity profile is fit
with a Gaussian profile.  For the template spectra used in both
methods we chose a K4.5 supergiant spectrum from the 3D stellar
library \citep{SRE1999}.  While the FCQ method is rather insensitive
to template mismatch, the results from the direct fitting method
depend very much on the choice of the stellar template spectrum.  We
find that the best choice for the type of the stellar template is a
late K or an early M supergiant (see also \S\ref{Sec:Continuum}).  The
errors of both the FCQ and the direct fitting method were determined
by modeling of galaxy spectra from stellar spectra with different
noise levels and trying to recover the input spectrum.  With the
direct fitting method a $\chi^2_{\mathrm{red}}$-analysis was used as
an independent way to determine the fit error.

\subsection{Stellar Mean Velocity field: Rotation and Dynamical Mass}
\label{Sec:Rotation}\label{Sec:DynamicalMass}
Because of the sharp edged profile of the CO absorption bandhead we
were able to determine the stellar velocity field for the entire
central region of \objectname[]{NGC~6240} with the direct fitting
method.  The resulting velocity field is shown in contour form in
Figure \ref{Fig:Velocityfield} superposed on the NICMOS K-band HST
archival image (PI: N. Scoville, proposal~ID: 7219).  Velocity
gradients across both nuclei are clearly recognizable.  The velocity
error at the position of the southern nucleus is $\pm18\kms$ and
increases to $\pm50\kms$ at a distance of $0\farcs6$; for the northern
nucleus the value is $\pm60\kms$ and $\pm85\kms$ at a distance of
$0\farcs5$.  Along the edge of the velocity field shown in Figure
\ref{Fig:Velocityfield} the uncertainty is $\pm150\kms$.

The southern nucleus shows a velocity gradient at a position angle of
$34\arcdeg$ west of the north-south axis with the south-east of the
nucleus being redshifted.  The velocity gradient of the northern
nucleus has a position angle of $41\arcdeg$ east of the north-south
axis.  The north-east of the northern nucleus is redshifted with
respect to the center.  The maximum relative velocity shift from the
center along the velocity gradient within $1\arcsec$ is $\pm165\kms$
and $\pm170\kms$ for the southern and northern nucleus, respectively.
The agreement of the K-band morphology from NICMOS with the
kinematical morphology of the stellar velocity field is remarkable.
The two nuclei are elongated along the velocity gradient in the
velocity field, indicating individual inclined rotating disks in the
nuclei.  The northern nucleus is redshifted with respect to the
southern nucleus by $\approx50\kms$.  This value is smaller than the
value measured in the \Brg\ line and measured by \citet{FRI1985} but
consistent within the uncertainties of our measurement.

The stellar velocity is surprisingly different compared to the
velocity field of the molecular gas.  The gas forms a rotating disk
\emph{between} the two nuclei \citep{TAC1999} with a axis of rotation
not aligned with either of the stellar disks in the nuclei.  This
implys a decoupling of the stellar motions from the gas motions, most
likely caused by the tidal forces of the interaction.

The dynamical mass of the \objectname[]{NGC~6240} nuclei can be
determined from the stellar velocity field.  Under the assumption that
the stars move on circular orbits around the center of each nucleus,
the dynamical mass within the radius $R$ for a inclination corrected
rotation velocity $v_{\mathrm{rot}}$ is given by
\begin{equation}
\label{Equ:DynamicalMass}
{\cal M}_{\mathrm{dyn}}=2.3\times10^2\,\left(\frac{R}{\mathrm{pc}}\right)
  \cdot\left(\frac{v_{\mathrm{rot}}}{\mathrm{km\,s^{-1}}}\right)^2\Msol.
\end{equation}
The measured recessional velocities for the two nuclei are shown in
Figure \ref{Fig:Rotationcurve}.  We fitted a model rotation curve to
the data with a solid body ($v\propto R$) part for radii $R\le100\pc$
and a flat part at a rotation velocity $v_{\mathrm{rot}}$ for larger
radii.  The model disk was inclined towards the line of sight by an
angle of $45\arcdeg$ and $60\arcdeg$ for the northern and southern
nucleus, respectively.  The inclination angles were determined by
fitting ellipses to the isophotes of the NICMOS H-band image.  The
model disk was convolved with a Gaussian profile with $0\farcs8$ FWHM,
corresponding to the average observing conditions.  For the assumed
inclinations the best fit of the rotation curve is achieved with
$v_{\mathrm{rot}}=270\pm90\kms$ for the southern nucleus and
$v_{\mathrm{rot}}=360\pm195\kms$ for the northern nucleus.  Even for
an inclination of $90\arcdeg$ the rotation velocities are
$180\pm90\kms$ and $200\pm155\kms$, respectively.  Assuming an extreme
inclination angle, the dynamical mass within $500\pc$ of the northern
nucleus is $2.4\times10^9\Msol$, and $1.9\times10^9\Msol$ for the
southern nucleus.  Such large masses can only be explained if the
observed K-band nuclei are indeed true nuclei of the two merging
progenitor galaxies.  Given the likely inclinations indicated by the
NICMOS images the true masses are likely even factors $\gtrsim2$
higher.  Table \ref{Tab:Dynamicalmass} lists the derived dynamical
masses of both nuclei for an inclination of $45\arcdeg$ and
$60\arcdeg$ for the northern and southern nucleus, respectively.

We also list in Table \ref{Tab:Dynamicalmass} the ratio of dynamical
mass to stellar mass content of the starburst derived from the
light-to-mass ratio.  The dynamical mass exceeds the visible mass by
about an order of magnitude.  The missing mass cannot be explained by
a contribution of molecular gas at the position of the nuclei.  The
molecular gas is concentrated between the nuclei with a relatively
small mass at the position of the nuclei \citep{TAC1999}.  From the
millimeter \COtwoone\ line flux within an $1\arcsec$ diameter aperture
on the nuclei an upper limit for the cold molecular gas mass in the
nuclei can be determined.  For the northern and southern nucleus this
upper limit is $0.2$ and $0.25$, respectively, of the gas mass within
the central $2\arcsec$ of the CO-disk.  Taking $2\times10^9\Msol$ for
the mass of the gas concentration \citep{TAC1999} the contribution
from the cold molecular gas to the total mass of the nuclei is
$4-5\times10^8\Msol$, much smaller than the missing mass.  The
contribution from a dark matter halo to the bulge masses cannot be
determined from our data.  In normal spiral galaxies this contribution
is negligible.  The only remaining likely source of missing mass is in
the old stellar population of the progenitor galaxies themselves.  The
number of faint stars could add up to a significant mass, yet their
luminosity would be negligible compared to the red supergiants of the
recent starburst.  We conclude that the two K-band peaks are indeed
the central bulges of the progenitor galaxies and that these bulges
are fairly massive, in accordance with the prediction of
\citet{MIH1996} and \citet{BAR1996}.

\subsection{Stellar Velocity Dispersion: Mass Concentration between
the Nuclei}
While the sharp edge of the \COtwozero\ bandhead allows a redshift
determination even for data with a low signal-to-noise ratio, we
cannot determine the velocity dispersion from such data.  We have,
therefore, completed the analysis of the spatial variation of the
velocity dispersion with integrated spectra for several apertures
centered in the nuclear region of \objectname[]{NGC~6240}.  To
determine the true velocity dispersion we first subtracted the stellar
rotation component as derived above.  Within the selected apertures we
corrected the data cube for the stellar rotational velocity field by
shifting the spectra of all spatial pixels to the same restframe.
Figure \ref{Fig:Velocitydispersion} shows the distribution of the
velocity dispersion over the nuclear region of
\objectname[]{NGC~6240}.  We show the calculated velocity dispersion
for four apertures: two centered on the nuclei; one centered between
the nuclei; and one south of the southern nucleus.  The error bars
along the declination axis denote the width of the apertures over
which the spectrum was integrated.  The aperture width in right
ascension is $0.8\arcsec$.  Figure \ref{Fig:Velocitydispersion} shows
that the velocity dispersion peaks \emph{between} the two nuclei at
$\sigma=276\pm51\kms$, while the velocity dispersion at the northern
and southern nucleus is $\sigma=174\pm54\kms$ and
$\sigma=236\pm24\kms$, respectively.  Because this result is derived
from the velocity field corrected data cube, the broad velocity
profile cannot be due to the overlapping of two velocity profiles from
components separated by scales larger than the spatial resolution of
our observations, that is $350$ to $400\pc$.

Replacing $v_{\mathrm{rot}}$ with the velocity dispersion $\sigma$,
equation (\ref{Equ:DynamicalMass}) can also be used to estimate the
mass ${\cal M}=2.3\times10^2\cdot
R_\mathrm{pc}\cdot\sigma_\mathrm{km\,s^{-1}}^2\Msol$ within the radius
$R$.  This is a very simplistic form of the Jeans equation
\citep{BIN1987} where only the isotropic part is considered and
numerical factors from the space distribution of the stars and the
spatial dependency of rotation and velocity dispersion are neglected
(thus resulting likely in an under-estimate of the dynamical mass).
Following this argument the peak of the velocity dispersion indicates
a mass concentration between the two nuclei.  We get masses for the
northern and southern nucleus of $0.8\pm0.5\times10^9\Msol$ and
$1.8\pm0.4\times10^9\Msol$, respectively.  Within the uncertainties
this is consistent with the dynamical masses derived above.  The mass
between the nuclei is $2.1\pm0.8\times10^9\Msol$.  The value of the
masses determined in this way are somewhat questionable, since the
Jeans equation applies to relaxed systems, a requirement that is not
fulfilled in \objectname[]{NGC~6240}.  At the peak of the stellar
velocity dispersion little stellar continuum is detected.  The low
continuum flux between the nuclei cannot be explained by extinction
which, although it peaks between the nuclei, is too low to explain a
hidden concentration of stars.  On the other hand, \citet{TAC1999}
find a concentration of cold molecular gas of
$\approx2\times10^9\Msol$ centered within $1\kpc$ between the nuclei.
It is thus very likely that the peak of the stellar velocity
dispersion is caused by this molecular gas concentration.  Our
extinction and \Htwo\ maps indicate also that there is a corresponding
peak of dust and hot $\mathrm{H_2}$.

\section{Molecular Gas}
\subsection{\mathversion{bold}Shock Excited $\mathrm{H_2}$ Emission}
\label{Sec:H2}
From an analysis of the near-infrared $\mathrm{H_2}$ line ratios
\citet{VWE1996} and \citet{SUG1997} conclude that $\mathrm{H_2}$ is
thermally excited with an excitation temperature of
$\approx2000\mathrm{\,K}$.  The four $\mathrm{H_2}$ lines we observed
also fit this result.  With the Infrared Space Observatory (ISO) pure
rotational $\mathrm{H_2}$ lines were detected for the first time in
\objectname[]{NGC~6240} \citep{LUT1996,EGA1998,RIG1999}.  The analysis
of their line ratios yields an excitation temperature
$<400\mathrm{\,K}$.  \citet{VWE1996}, \citet{SUG1997} and
\citet{EGA1998} all find a slow continuous shock (C-shock) as the best
fitting model for the $\mathrm{H_2}$ excitation.  Fast shocks would
lead to a discontinuous or jump shock (J-shock).  Due to the
discontinuity J-shocks ionise the interstellar medium and hence
emission from ionised elements is expected.  The low \Brg\ to \Htwo\
line ratio therefore indicates strongly that the shock is
non-dissociative.  The agreement of the \Brg\ and continuum morphology
suggests the starburst as the source of the \Brg\ emission.

In the C-shock models the shock has a velocity $\lesssim40\kms$, yet,
the observed $\mathrm{H_2}$ line widths are $\Delta v\approx550\kms$.
The large line width are most likely a superposition of several
narrower $\mathrm{H_2}$ lines with different radial velocities along
the line-of-sight.  If we consider the distribution of molecular gas
in form of dense clouds within a less dense intercloud medium we can
explain the slow shock speed.  The collision between molecular clouds
occurs under a variety of angles and the shock speed therefore is only
the projected radial velocity difference.  But even for a direct
collision, the shock speed in the denser cloud is smaller than in the
lower density intercloud medium.  An originally fast shock propagating
in the intercloud medium will propagate with a slower shock speed when
it enters a medium of higher density like a molecular cloud.  It is
these secondary slow shocks in the dense clouds that we are observing
in the near-infrared ro-vibrational $\mathrm{H_2}$ lines.

\subsection{Cold and Hot Molecular Gas}
\label{Sec:H2+CO}
The gas motions in the central region of \objectname[]{NGC~6240} are
highly turbulent.  \citet{TAC1999} find \COtwoone\ line width of up to
$400\kms$ FWHM (and $1000\kms$ FWZP) and their channel maps show
filaments extending from a central CO-disk out to $2\kpc$.

A comparison of Figure \ref{Fig:Channel Maps} with the channel maps of
the \COtwoone\ line \citep[][Figure 2]{TAC1999} indicates that the
cold molecular gas (\COtwoone) and the hot, shock-heated
$\mathrm{H_2}$ have a very similar overall morphology.  Local
differences between the \COtwoone\ and \Htwo\ channel maps are
apparent.  The most obvious difference is in the central region
between the nuclei.  The hot gas is more extended than the cold gas
and a disk structure as seen in the \COtwoone\ line is not visible.
In addition, the \Htwo\ line is $150\kms$ to $250\kms$ broader than
the \COtwoone\ line.  Both effects can be explained with a different
distribution of the cold and hot gas.  Roughly half of the cold gas
has already settled in the center between the nuclei, while the
shock-heated interstellar medium of the two galaxies is still more
extended.  The $\mathrm{H_2}$ emission is excited on the surfaces of
molecular clouds while the \COtwoone\ emission originates more in the
volume of the molecular cloud.  Because we sample with the
$\mathrm{H_2}$ emission a larger volume than with the \COtwoone\
emission, we also sample a larger range in radial velocities and hence
observe larger linewidths for the $\mathrm{H_2}$ lines.  The larger
line width of the \Htwo\ lines are not symmetric about the peak of the
\COtwoone\ lines but are blueshifted by $\approx150\kms$.

In the structure outside the nuclear region of \objectname[]{NGC~6240}
the \Htwo\ and \COtwoone\ emission is dominated by a filamentary
structure.  The filaments in the south-east and south-west are the
most prominent and show similar morphology and kinematics in the
\COtwoone\ and \Htwo\ line.  Also the weaker filaments north of the
nuclei show this similarity.  They are most likely gas flows towards
the centers of the nuclei.

\begin{figure*}
\includegraphics[width=\textwidth]{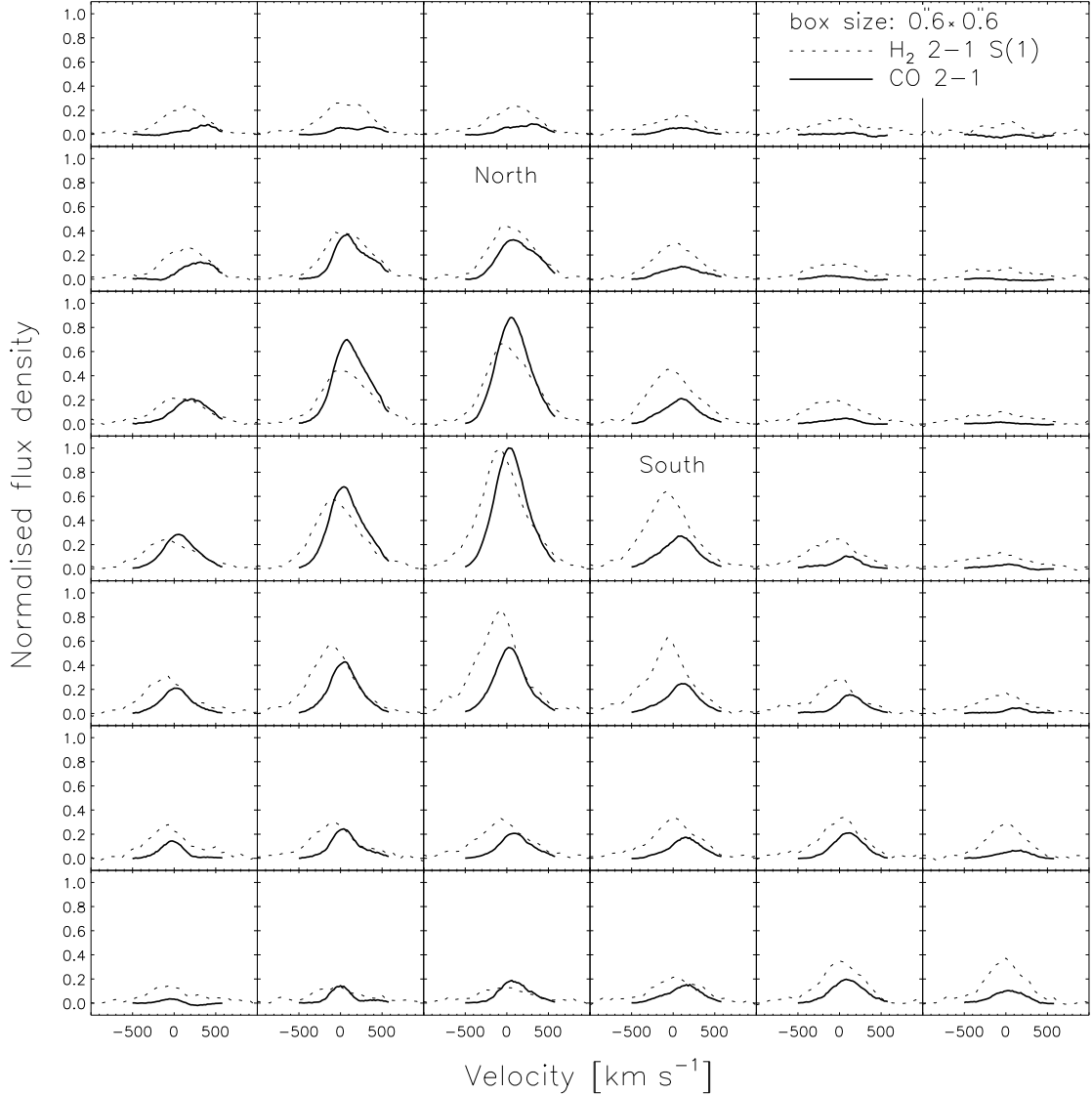}
\figcaption{
Line-of-sight velocity profiles of the \Htwo\ and \COtwoone\
molecular emission lines at a velocity resolution of $140\kms$.  The
dashed line represents the \Htwo\ profile, the solid line the
\COtwoone\ profile.  The aperture size over which the profiles were
integrated is $0\farcs6\times0\farcs6$.  The profiles at the position
of the two nuclei are indicated by the labels ``North'' and ``South''.
\label{Fig:H2&CO Profiles}}
\end{figure*}

\begin{figure}
\includegraphics[width=\columnwidth]{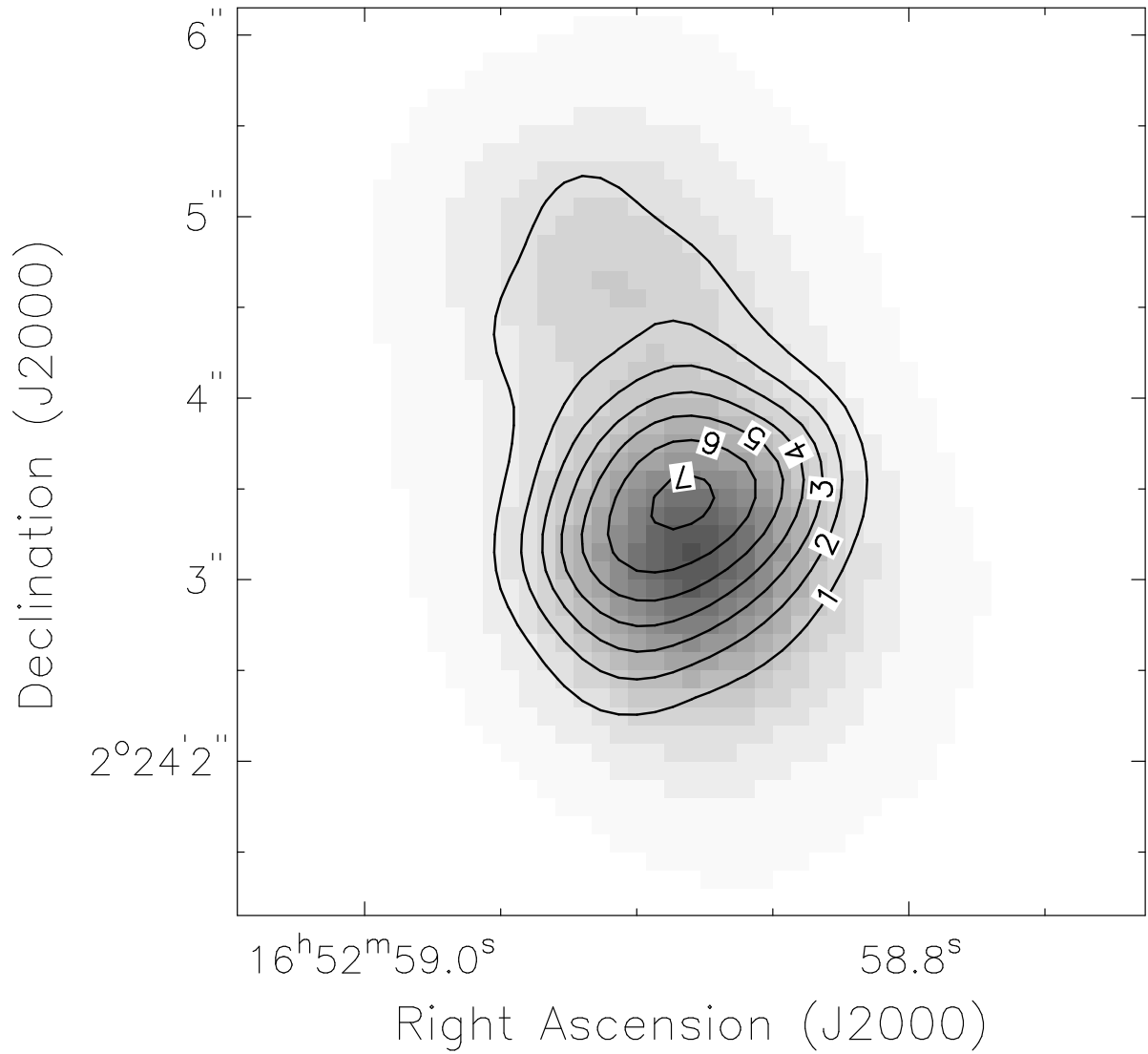}
\figcaption{
Extinction map of NGC~6240 in contours over a grayscale
representation of the K-band map.  The numbers give $\mathrm{A_V}$ for
the foreground screen model.
\label{Fig:Extinction}}
\end{figure}

\begin{figure}
\includegraphics[width=\columnwidth]{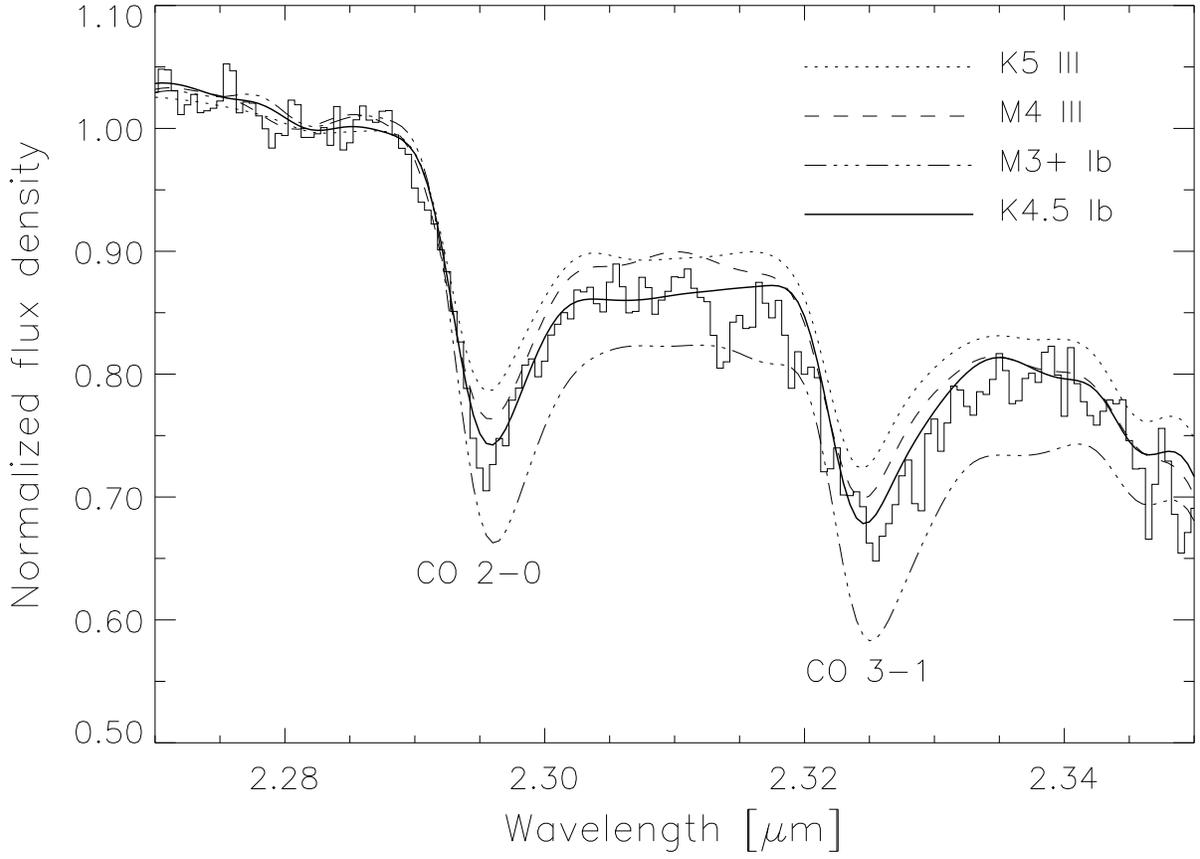}
\figcaption{
Spectrum of the southern nucleus of NGC~6240 together with spectra of
four giant and supergiant stars.  For convenient comparison the
spectrum of NGC~6240 was blueshifted into the restframe of the stellar
spectra.  The stellar spectra were convolved with a Gaussian profile
with a FWHM corresponding to $600\kms$.
\label{Fig:Supergiants}}
\end{figure}

\begin{figure}
\includegraphics[height=0.75\textwidth]{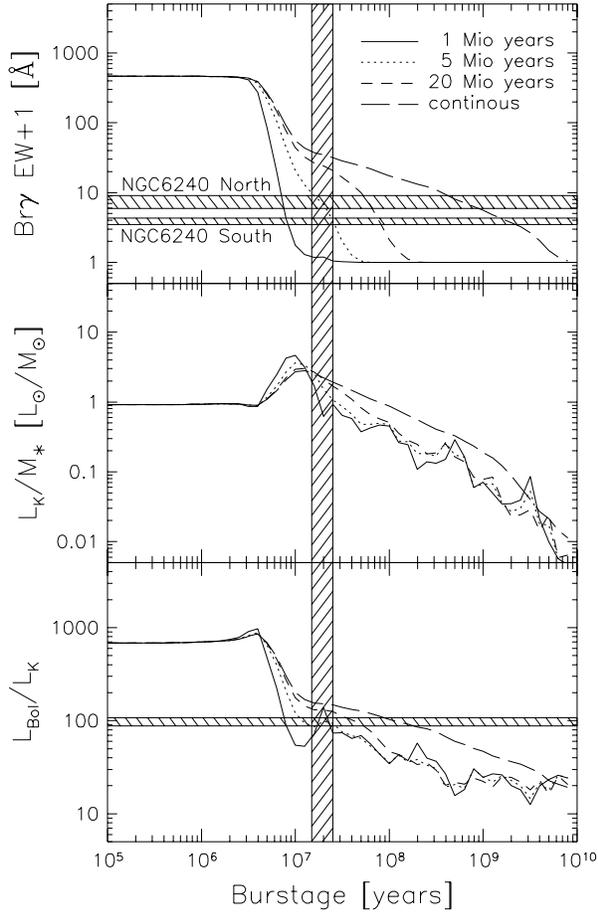}
\figcaption{
Results from starburst simulations as a function of starburst age.
Shown are four starburst scenarios with continuous star formation and
starburst durations of $1$, $5$ and $20$ million years.  The vertical
hatched bar denotes the age range from $15$ to $20$ million years, the
age of a red supergiant.
Top: \Brg\ equivalent width as a function of starburst age.  The
measured equivalent widths of the southern and northern nucleus are
denoted by horizontal hatched bars.
Middle: K-band luminosity to stellar mass ratio as a function of
starburst age.
Bottom: Bolometric luminosity to K-band luminosity ratio as a function
of starburst age.  The horizontal hatched bar denotes the value for
NGC~6240 in a $5\arcsec$ diameter circular aperture.
\label{Fig:Starburst}}
\end{figure}

\begin{figure}
\includegraphics[width=\columnwidth]{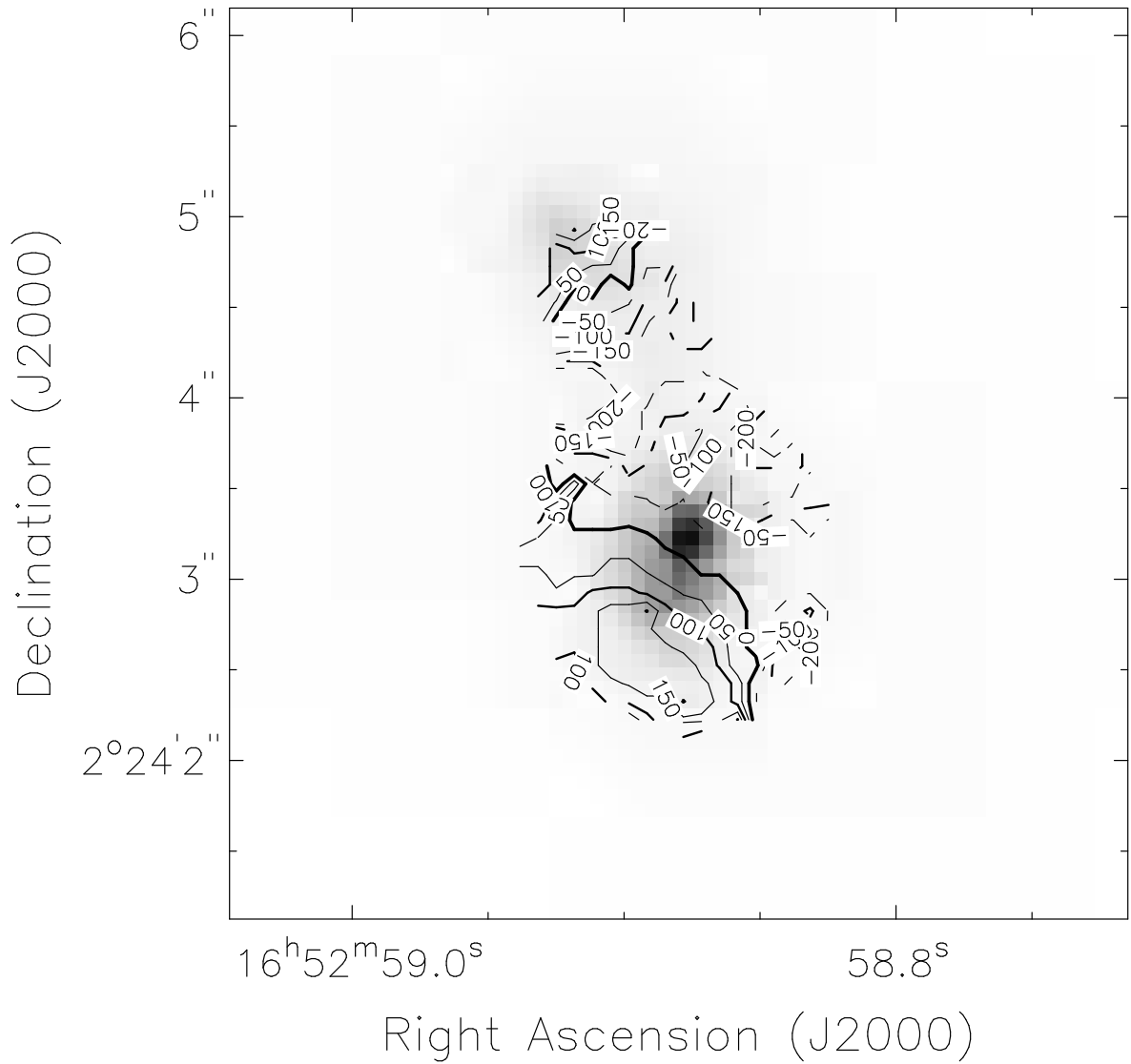}
\figcaption{
Stellar velocity field of NGC~6240 in contours over a gray scale
representation of the NICMOS K-band map.  The contours are spaced
$50\kms$, solid contours represent positive, redshifted velocities,
dashed contours represent negative, blueshifted velocities.  The
systemic velocity used is as $7300\kms$.
\label{Fig:Velocityfield}}
\end{figure}

\begin{figure}
\includegraphics[width=\columnwidth]{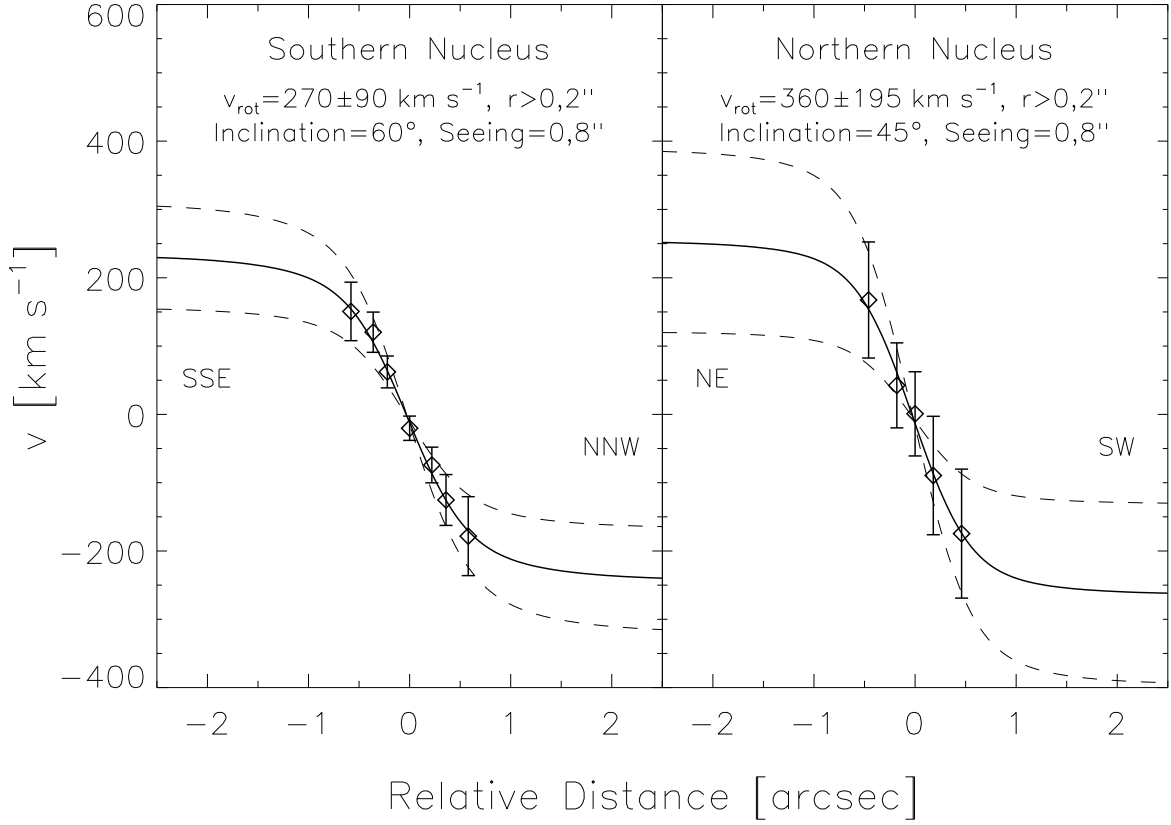}
\figcaption{
Measured and model rotation curves for the northern (right panel) and
southern (left panel) nucleus of NGC~6240.  Open diamonds represent
the measured rotation velocities, the solid line is the best fitting
model rotation curve, and the dashed lines denote the $1\sigma$ error
of the model rotation curves.
\label{Fig:Rotationcurve}}
\end{figure}

\begin{figure}
\includegraphics[width=\columnwidth]{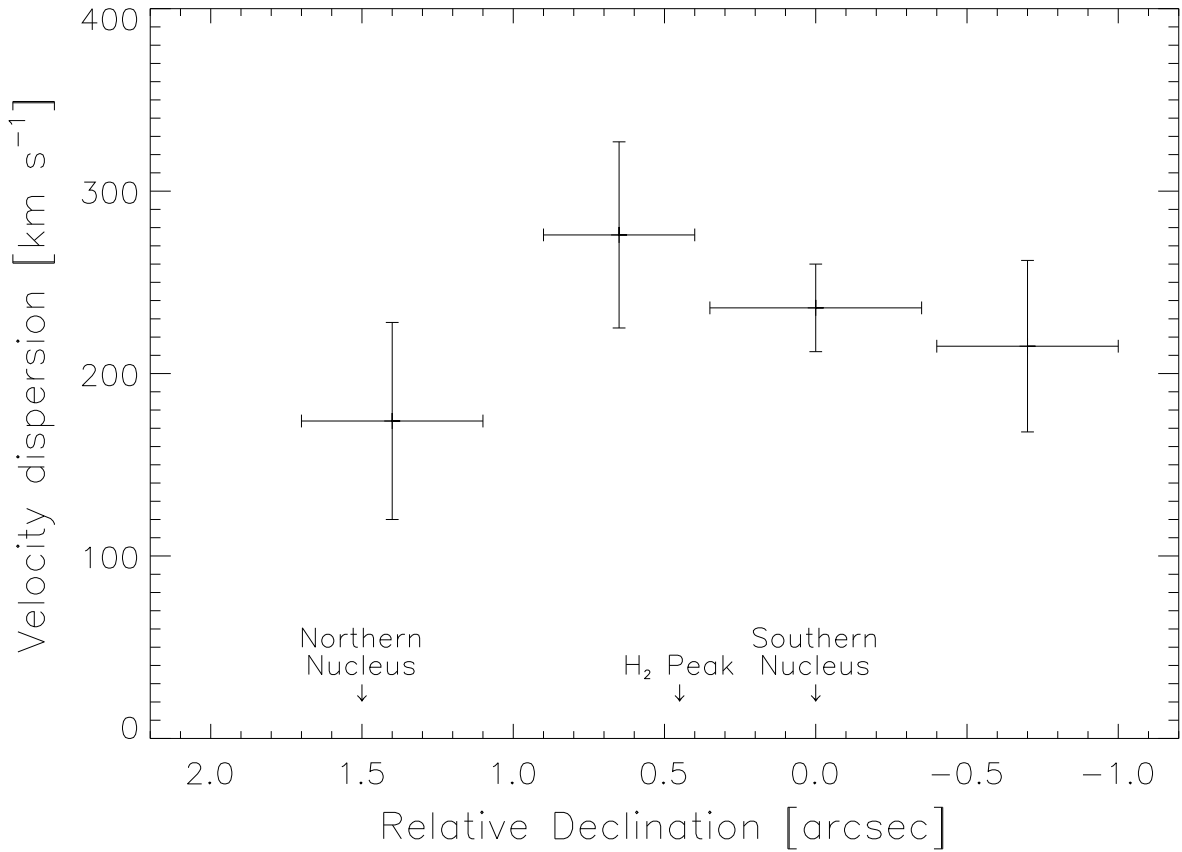}
\figcaption{
Stellar velocity dispersion of NGC~6240 along a north-south direction.
The error bars along the abscissa represent the width of the aperture,
over which the spectra were integrated.  Velocity error bars are
$1\sigma$ error bars.
\label{Fig:Velocitydispersion}}
\end{figure}

\begin{figure*}
\includegraphics[width=\textwidth]{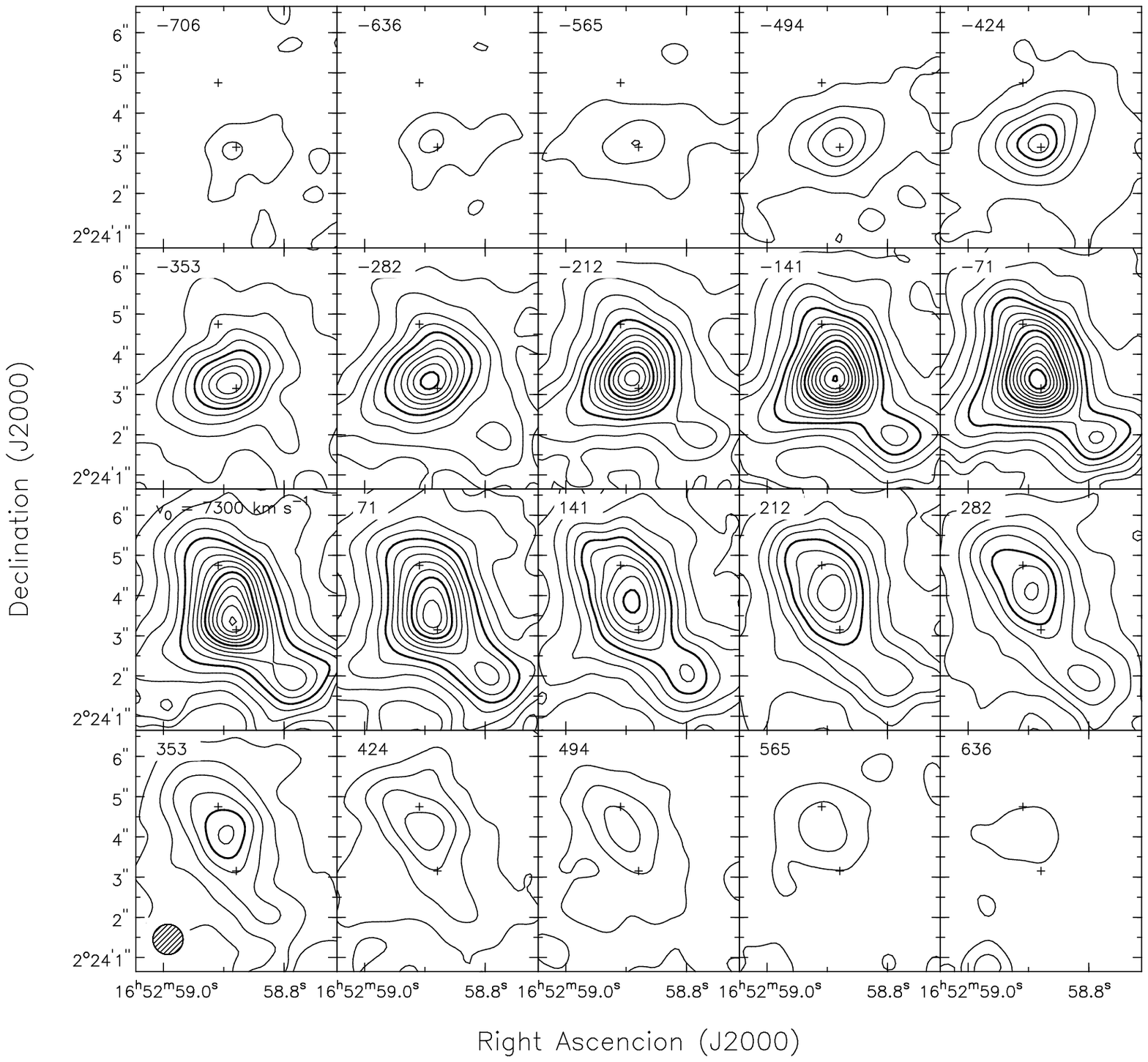}
\figcaption{
Channel Maps of the \Htwo\ emission line at velocity intervals of
$70\kms$.  Contour levels start at
$5\times10^{-6}\mathrm{\,erg\,s^{-1}\,cm^{-2}\,st^{-1}}$ in intervals
of $5\times10^{-6}\mathrm{\,erg\,s^{-1}\,cm^{-2}\,st^{-1}}$.  In the
upper left corner of each panel the radial velocity with respect to
$7300\kms$ is plotted.  Spatial resolution is
$0\farcs75\times0\farcs75$.
\label{Fig:Channel Maps}}
\end{figure*}

A comparison of the velocity profiles of the \COtwoone\ and \Htwo\
lines is shown in Figure \ref{Fig:H2&CO Profiles}.  From the
line-of-sight velocity profiles in Figure \ref{Fig:H2&CO Profiles}
several kinematical components of the \Htwo\ line are visible.  They
vary over the field, especially at the positions where the filaments
extend from the \Htwo\ emission peak.  North-east of the northern
nucleus two components are recognizable with a velocity difference of
$\approx250\kms$.  The filament in the south-west shows also two major
components with a velocity difference of $\approx200\kms$.  The \Htwo\
line exhibits this multi-component character over the entire emission
region of \objectname[]{NGC~6240}.

The radial velocity difference between northern and southern nucleus
is $\approx150\kms$ in the \Htwo\ line and $\approx100\kms$ in the
\COtwoone\ line.  This is similar to the radial velocity difference of
the nuclei measured in the \Brg\ line.  Globally the molecular gas
appears to follow the motion of the two nuclei.  As with the cold gas,
a rotation of the hot gas seems the best explanation for the velocity
gradient of the \Htwo\ line.  The velocity gradient peaks between the
nuclei at the position of the \COtwoone\ emission peak.  At this
position the velocity dispersion of the \Htwo\ emission also reaches
its maximum of $240\kms$.

The central disk of cold molecular gas has a rotation axis which is
not correlated with the rotation axis of either nucleus.  Rather the
sense of the rotation corresponds to the relative motions of the
nuclei. The CO-disk seems to retain the memory of the orbital history
of the interaction and is independent of the gravitational forces
exerted by the nuclei.

\section{Interaction and Merging}
We have shown that \objectname[]{NGC~6240} is a merging system of two
galaxies with massive bulges.  In the following sections results from
our observations and published results are used to draw a more
detailed picture of the interaction in \objectname[]{NGC~6240}.

\subsection{Prograde Encounter?}
\label{Sec:Prograde?}
The rotation velocities for both nuclei, derived in
\S\ref{Sec:Rotation}, are much larger than the relative velocity
difference between the two nuclei.  Assuming circular orbits for the
nuclei the small radial velocity difference and projected distance
between the nuclei yields a small projected orbital angular momentum
of the nuclei.  This means that the two galaxies either have lost a
major fraction of their original orbital angular momentum in the
interaction, or that the true orbital angular momentum is much larger
than the projected value.  A larger true orbital angular momentum
means that the actual distance between the two nuclei is larger than
the measured, projected separation.  On the other hand, during the
interaction of two galaxies their orbits can vary and the nuclei might
rather be on radial than on circular orbits.  In this case their
orbital momentum could be still large.  The exact trajectories of the
galaxies depend on the initial conditions of the interaction.

\begin{figure*}
\includegraphics[width=\textwidth]{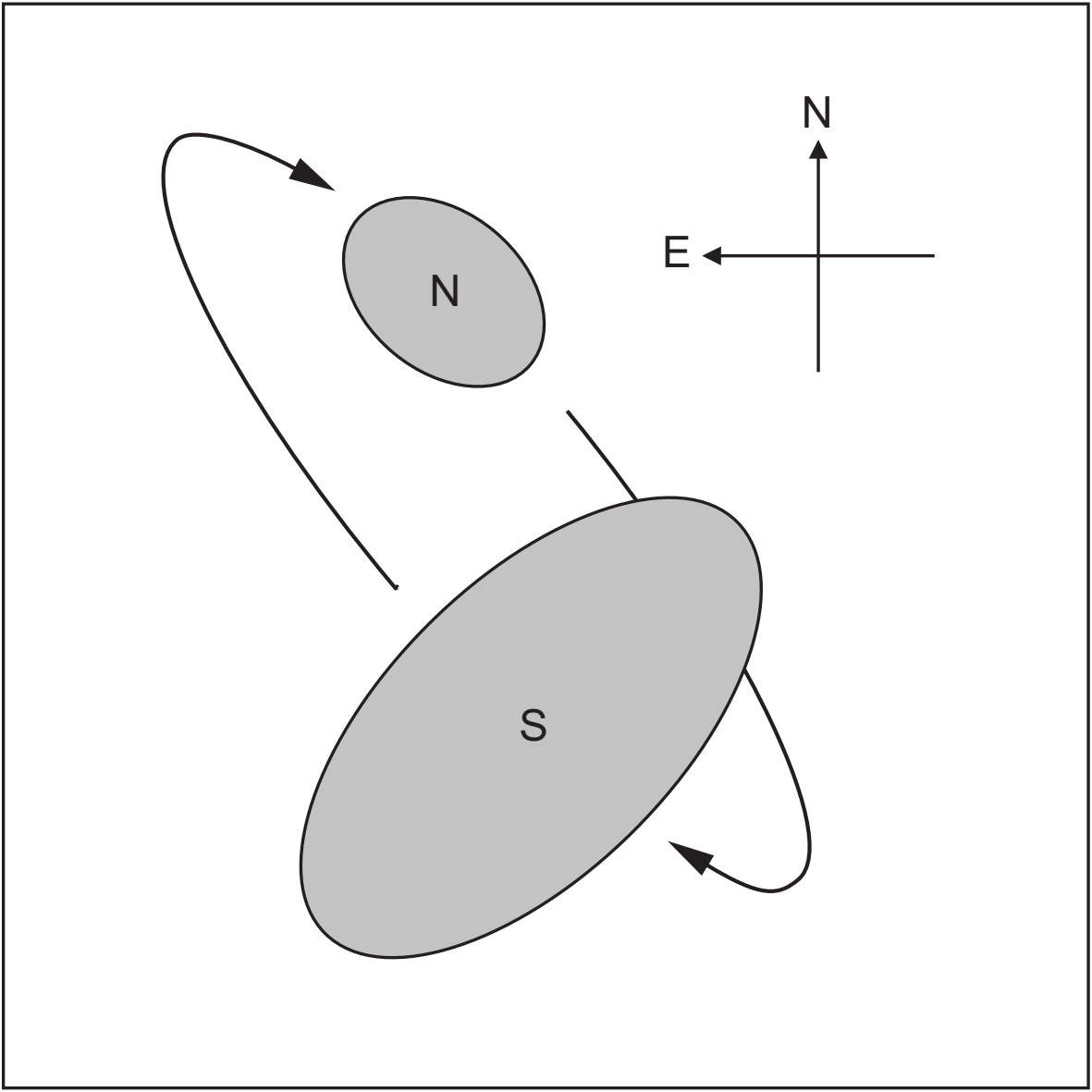}
\figcaption{
Schematic view of the assumed collision geometry in NGC6240.  For the
drawn geometry the receding northern nucleus is further away than the
southern nucleus.
\label{Fig:Collision}}
\end{figure*}

Because we can only observe projected angular momenta, the exact
orientation of the disks with respect to the orbital plane cannot be
determined.  A schematic view of the collision geometry is shown in
Figure \ref{Fig:Collision}.  The projected spin angular momentum of
the northern nucleus is almost parallel to the systems projected
orbital angular momentum, while the projected spin angular momentum of
the southern nucleus is nearly normal to it.  If the projected angular
momenta are the true angular momenta, one partner in the collision,
now seen as the northern nucleus, was subject to a prograde encounter,
while the other partner is inclined with respect to the orbital plane.
For the motion of the two nuclei around each other we assume a
circular orbit, whose position angle and inclination is equal to the
position angle and the inclination of the rotating CO-disk between the
nuclei \citep{TAC1999}.  Under this assumption and with a radial
velocity difference of $150\kms$ the true distance between the two
nuclei can be calculated.  With a position angle of the CO-disk of
$40\arcdeg$ and an inclination of $75\arcdeg$ the true separation of
the two nuclei is $1.4\kpc$, their orbital velocity is $155\kms$ with
an orbital period of $27$ million years.  Assuming a simple two-body
problem with equal mass of $2\times10^9\Msol$, this orbit is unstable
with the gravitational force being $\slantfrac{1}{8}$ of the
centrifugal force.  This means that the two nuclei will separate in
the future and probably are already past the pericenter.  Whether the
system is bound can be estimated from the escape velocity. We compute
an escape velocity of $110\kms$, which is of the same order as the
radial velocity difference of the two nuclei.  From the radial
velocity difference the system seems to be unbound. However, if the
nuclei separate again they will see not only the mass of the other
nucleus but also the halo of the other galaxy.  Conversion of orbital
angular momentum into spin angular momentum in the halo leads to
gravitational braking, the galaxies reach their apocenter and fall
back together again \citep{BAR1996,MIH1996}.  It seems therefore
likely, that \objectname[]{NGC~6240} has recently undergone an
encounter and is awaiting its next encounter.

\subsection{Tidal tails}
The first indication that \objectname[]{NGC~6240} might be an
interacting galaxy system was its disturbed morphology with loops,
branches and arms.  \citet{TOO1972} showed that such tails can be
formed by tidal forces in an interaction of two galaxies.  Tidal
forces during the interaction pull stars from the galaxies into
regions at a distance of up to several tens of kiloparsec from the
center of the galaxies.  The formation of tidal tails is therefore an
efficient way to dissipate orbital angular momentum of the interacting
galaxies thus allowing the galaxies to come closer to each other and
finally merge.  That the tails in \objectname[]{NGC~6240} are
dominated by stellar light can be seen from broad and narrow band
imaging.  The tidal tails visible in the R-band image of
\citet{ARM1990} are not as prominent in their narrow band image of
$\mathrm{H\alpha}+$[\ion{N}{2}] which are the most dominant emission
lines in this wavelength band.  The formation of tidal tails is
favoured if at least one partner in the collision is subject to a
prograde encounter.  The spin and orbital angular momenta of the
nuclei suggest a prograde encounter for what is now the northern
nucleus.  But not only the geometry of the collision is important for
the formation of tidal tails.  \citet{SPR1998} and \citet{DUB1999}
show that the mass density distribution of the interacting galaxies is
an important parameter in the formation of tidal tails.  Galaxies with
more massive halos tend to form less sharp and shorter tidal tails.
The tidal tails in \objectname[]{NGC~6240} are not as ``crisp'' as in
\objectname[]{NGC~4038/9} (``The Antennae'') or in
\objectname[]{NGC~4676} (``The Mice''), indicating the collision of
two massive spiral galaxies.  Because the parameter space of galaxy
interaction models is too large, it is impossible to derive a detailed
history of \objectname[]{NGC~6240} from the morphology of its tidal
tails.  However, simulations of interacting and merging galaxies show,
that tidal tails can only be formed \emph{after} the first encounter
of the galaxies.  Because we see tidal tails in
\objectname[]{NGC~6240} which are quite long, it must have undergone
at least one encounter.  The tidal tails of \objectname[]{NGC~6240}
were most likely created in the first encounter.  How many passages
\objectname[]{NGC~6240} has undergone we cannot determine from the
tidal tails.  But because the two nuclei are still distinct from each
other, it seems that only few encounters can have happened and that
\objectname[]{NGC~6240} is in a rather early merger phase.

\citet{DOY1994} showed from K-band photometry that the overall
brightness distribution of \objectname[]{NGC~6240} is very similar to
that of an elliptical galaxy.  They concluded that
\objectname[]{NGC~6240} is in an advanced merger state and in the
process of forming an elliptical galaxy.  We believe that this
evidence is rather weak and the conclusion should be treated with
caution.
\begin{enumerate}
\item The near-infrared light of the nuclei comes from supergiants and
not from the old stars representing the overall stellar population.
Population and $L/{\cal M}$ changes with radius are rather likely.
\item While a $r^{1/4}$-law or a King-profile fits the data it is not
unique to elliptical galaxies.
\item The stellar dynamics definitely shows that the two galaxies are
still independent entities and have not yet merged.
\end{enumerate}

However, the small separation of the nuclei and the decoupling of the
gas from the stars in \objectname[]{NGC~6240} strongly suggest that
the nuclei will merge.  But as the formation of tidal tails depends on
the mass density distribution of the interacting galaxies the merging
time also varies with the mass density distribution.  Galaxies with
extended massive halos have longer merging times than galaxies with
compact low mass halos.  Only in the final merging phase violent
relaxation leads to a strong and rapid dynamical evolution also on
global scales \citep{MIH1998}.

What type of galaxy the merger remnant will resemble we cannot predict
from the data.  However, \objectname[]{NGC~6240} might form an
elliptical galaxy of which it shows already signatures.  The stellar
velocity dispersion in the nuclei of \objectname[]{NGC~6240} is
typical for an elliptical galaxy.  Outside the nuclear region the
K-band light can be fit by a $r^{1/4}$-law which also approximates the
surface brightness profile of merger remnants in numerical simulations
of galaxy mergers \citep{BAR1992}.

\subsection{Nuclear Starbursts}
Another hint at how many passages \objectname[]{NGC~6240} has
already undergone comes from the age and duration of the nuclear
starbursts.  In \S\ref{Sec:Starburst} we determined the age of the
starburst to be $15$ to $25$ million years.  Numerical simulations
predict that such nuclear starbursts in merging galaxies are
triggered by close passages of the two galaxies.  If this was true,
the last encounter happened $\approx20$ million years ago.  Because
the tidal tails are quite long and pronounced, the first encounter
must be well past and the last passage was probably at least the
second encounter.  The dynamical time-scale of the orbital motion of
the two nuclei is a few $\approx10$ million years and is of the same
order as the age of the starburst.  The extent of the superwind
($\approx5\kpc$) and its velocity ($\approx500\kms$) yields an age for
the superwind of $\approx10$ million years, similar to the lifetime of
a superwind \citep{HEC1990}.  If taken into account that the superwind
sets in roughly $10$ million years after the starburst, the superwind
and the K-band continuum can be explained by the same starburst.  The
duration of the starburst is much shorter than the age of the
starburst and can be explained by a negative feedback effects of the
starburst through supernovae and stellar winds.  Numerical simulation
by \citet{MIH1996} predict an episodic star formation during the
merging of two galaxies.  Gas streaming into the central regions gets
compressed and triggers star formation.  Depending on the existence of
massive bulges one major starburst takes place either after the first
encounter or during the final merging of the two galaxies.  Because
\objectname[]{NGC~6240} is clearly not in the stage of final merging,
the starburst we see has likely been triggered in an early encounter.

\subsection{Gas Concentration between the Nuclei}
The presence of a self-gravitating cold molecular gas concentration
between the nuclei of \objectname[]{NGC~6240} is not expected from
numerical simulations of galaxy mergers \citep{BAR1996,MIH1996}.
These simulations predict that the highly dissipative gas follows the
stellar distribution.  Gas disks form around the individual galaxy
nuclei and get more compact as the merger advances.  From the point of
view of these simulations it would appear as if
\objectname[]{NGC~6240} has to be in a rare transient phase.  

However, \objectname[]{NGC~6240} is not the only galaxy with a gas
concentration between two nuclei.  \objectname[]{VV~114}
\citep{YUN1994}, \objectname[]{NGC~6090} \citep{GAO1998,BRY1999} and
\objectname[]{NGC~4038/9} \citep{STA1990} all have prominent
CO-emission peaks between the two nuclei.  In all cases the projected
distance between the nuclei ($6\kpc$ for \objectname[]{VV~114},
$3.5\kpc$ for \objectname[]{NGC~6090} and $7\kpc$ for
\objectname[]{NGC~4038/9}; $\mathrm{H_0}=75\kms\Mpc^{-1}$) is much
larger than the separation of $750\pc$ for \objectname[]{NGC~6240}.
\objectname[]{NGC~6240} is also an exception because very little
CO-emission is observed at the position of the nuclei.  In
\objectname[]{NGC4038/9} three clearly distinct gas concentrations are
detected: two at the position of the nuclei; and a third between the
nuclei, which contains roughly half of the total gas mass.
\objectname[]{NGC~6240}, too, has half of the molecular gas in a
central concentration between the nuclei but the other half is in
filaments extending from the central gas concentration.  In
\objectname[]{VV~114} the molecular gas is in a bar-like concentration
with tails extending from it.  \objectname[]{NGC~6090} shows a
ridge-like structure of molecular gas which covers also both nuclei.
\objectname[]{VV~114}, \objectname[]{NGC~6090} and
\objectname[]{NGC~4038/9} are all, mainly based on the large
separations, considered to be early mergers.  The small separation of
the two nuclei in \objectname[]{NGC~6240} could be interpreted as if
\objectname[]{NGC~6240} was in an advanced merger state.  However, to
derive the evolutionary state of merging galaxies only on the nuclear
separation can be misleading.  Due to projection effects and unknown
orbits of the galaxies the nuclear separation could be much larger.

\section{Conclusions}
From the high resolution near-infrared integral field spectroscopy of
\objectname[]{NGC~6240} we find:
\begin{itemize}
\item The K-band light of the two nuclei in \objectname[]{NGC~6240} is
dominated by red supergiants which must have been formed in a
starburst $15-25$ million years ago.  The duration of the star
formation is $\lesssim5$ million years, thus only a small fraction of
the starburst age.  The total mass of stars formed in the starburst is
${\cal M}_*=0.4-2\times10^8\Msol$.
\item The stars in the two K-band peaks exhibit fast rotation.  From
the stellar velocity field the dynamical mass of the nuclei is
determined to be ${\cal M}_{\mathrm{dyn}}=2-8\times10^9\Msol$ within
the central $500\pc$ with similar masses for both nuclei.  This
exceeds the mass of the most massive star forming regions by more than
$100$, implying that the two infrared emission peaks are the massive,
rotating bulges of two interacting and merging galaxies.
\item After correction for the rotation component the stellar velocity
dispersion peaks between the two nuclei at the emission peak of the
molecular gas.  The inferred mass concentration is a self-gravitating
concentration of molecular gas.
\item The $\mathrm{H_2}$ emission of \objectname[]{NGC~6240} peaks
between the two nuclei and is thermally excited in a slow, continuous
shock triggered by the collision of the two galaxies.  Molecular gas
streams to the centers of the galaxies where the turbulent motions
also give rise to shock excited $\mathrm{H_2}$ emission.
\item From the orbital and spin angular momenta of the interacting
galaxies it seems that one galaxy in the system is subject to a
prograde encounter.  Based on the morphology of the tidal tails, the
starburst scenario, the kinematics and the excitation of the
$\mathrm{H_2}$ emission, we argue that we observe
\objectname[]{NGC~6240} shortly after an early encounter which
triggered the observed starburst.
\end{itemize}

\acknowledgements The authors are grateful to the staff of the ESO
$2.2\,\mathrm{m}$ telescope and the Anglo-Australian Telescope for
their support during the observation.  We also would like to thank
S.~Mengel, J.~Gallimore, R.~Maiolino, A.~Krabbe and H.~Kroker for
their help setting up and operating the instruments 3D and ROGUE and
collecting the data.

\section*{}

\section*{}
\begin{deluxetable}{lcccc}
\tablenum{1}
\tablewidth{0pt}
\tablecaption{\label{Tab:Luminosity}
Broad and Narrow Band Luminosities of NGC~6240}
\tablehead{
\colhead{Wave Band} & \colhead{Aperture} & \colhead{Diameter} & 
  \colhead{$L$} & \colhead{$L_{der}$} \\
\colhead{(1)} & \colhead{(2)} &\colhead{(3)} &\colhead{(4)} &\colhead{(5)}}
\startdata
       &      & 1       & $5.8\times10^8$ & $6.8\times10^8$ \\
       & S & $1.5$ & $1.1\times10^9$ & $1.2\times10^9$ \\
       &      & 2       & $1.5\times10^9$ & $1.7\times10^9$ \\
K-Band & & & & \\
       &      & 1       & $2.2\times10^8$ & $2.3\times10^8$ \\
       & N & $1.5$ & $4.5\times10^8$ & $4.7\times10^8$ \\
       &      & 2       & $7.0\times10^8$ & $7.3\times10^8$ \\
\\
\mbox{$\mathrm{H_2}\:v\!=\!1\!-\!0\:\mathrm{S(1)}$} &
$\mathrm{H_2}$ & 4 & $2.9\times 10^7$ & $3.0\times 10^7$ \\
\mbox{$\mathrm{H_2}\:v\!=\!1\!-\!0\:\mathrm{S(0)}$} &
$\mathrm{H_2}$ & 4 & $9.3\times 10^6$ & $9.8\times 10^7$ \\
\mbox{$\mathrm{H_2}\:v\!=\!2\!-\!1\:\mathrm{S(1)}$} &
$\mathrm{H_2}$ & 4 & $4.0\times 10^6$ & $4.2\times 10^7$ \\
\mbox{$\mathrm{H_2}\:v\!=\!2\!-\!1\:\mathrm{S(2)}$} &
$\mathrm{H_2}$ & 4 & $1.8\times 10^6$ & $1.9\times 10^7$ \\
\\
 & S & $1.5$ & $9.5\times10^5$ & $1.1\times10^6$ \\
Brackett~$\gamma$ & N & $1.5$ & $8.0\times10^5$ & $8.3\times10^5$ \\
 & N3 & $1.5$ & $4.5\times10^5$ & $4.5\times10^5$ \\
\enddata
\tablecomments{%
Columns: (1) {Wave band;}
         (2) {Position of circular aperture: southern nucleus (S),
northern nucleus (N), \Htwo\ emission peak ($\mathrm{H_2}$) and radio 
peak N3 \citep[see][]{COL1994};}
         (3) {Diameter of circular aperture in seconds of arc;}
         (4) {Luminosity in $\Lsol$ ($\mathrm{H_0}=75\kms\Mpc^{-1}$);}
         (5) {Dereddened K-band luminosity in $\Lsol$.  Extinction
values are for a foreground screen model from \S\ref{Extinction}.}}
\end{deluxetable}

\begin{deluxetable}{cccccc}
\tablenum{2}
\tablewidth{0pt}
\tablecaption{\label{Tab:Dynamicalmass}
Stellar and Dynamical Masses of NGC~6240}
\tablehead{
\colhead{Nucleus} &   \colhead{${\cal M}_*$} & \colhead{$i$} &
  \colhead{$v_{\mathrm{rot}}$} & \colhead{$\Mdyn$} &
  \colhead{$\Mdyn/{\cal M}_*$} \\
\colhead{(1)} & \colhead{(2)} & \colhead{(3)} & 
  \colhead{(4)} & \colhead{(5)} & \colhead{(6)}}
\startdata
North & $0.4-1.2\times10^8$ & $45\arcdeg$ &
$360\pm195$ & $7.6^{+10}_{-5.9}\times 10^9$ & $\ge14$ \\
South & $0.8-2.3\times10^9$ & $60\arcdeg$ & 
$270\pm90$ & $4.3^{+3.3}_{-2.4}\times 10^9$ & $\ge8$ \\
\enddata
\tablecomments{%
Columns: (1) {Nucleus;}
         (1) {Stellar mass in $\Msol$ within $r<235\pc$;}
         (2) {Stellar mass in $\Msol$ within $r<235\pc$;}
         (3) {Inclination of stellar disks;}
         (4) {Rotation velocity in $\kms$;}
         (5) {Dynamical mass in $\Msol$ within $r<235\pc$;}
         (6) {Ratio of dynamical and stellar mass.}}
\end{deluxetable}

\end{document}